\documentstyle{aipproc} 
\newskip\humongous \humongous=0pt plus 1000pt minus 1000pt

\newif\ifdtup

\def\oldreffmt#1{\rlap{[#1]} \hbox to 2\parindent{}}

\def\figfmt#1{\rlap{Figure {#1}} \hbox to 1in{}}

\def\etal{\hbox{\it et al.}}

\def\VEV#1{\left\langle #1\right\rangle}

\def\slash#1{#1\!\!\!/\!\,\,}
\def\beq{\begin{equation}}
\def\eeq{\end{equation}}
\def\bea{\begin{eqnarray}}
\def\eea{\end{eqnarray}}
\def\half{\frac{1}{2}}

\def\bq{\begin{quote}}
\def\eq{\end{quote}}

\def\half{\frac{1}{2}}     

\def \gta {\mathrel{\vcenter
     {\hbox{$>$}\nointerlineskip\hbox{$\sim$}}}} 

\def \etal {{\it et al.}\ }
\relax
\def\a{\alpha}

\def\d{\delta}

\def\e{\epsilon}

\def\F{\Phi}

\def\L{\Lambda}

\def\q{\theta}

\def\fr{\frac}
\def\ba{\begin{array}}
\def\ea{\end{array}}
\def\bz{\begin{equation}}
\def\ez{\end{equation}}
\def\by{\begin{eqnarray}}
\def\ey{\end{eqnarray}}

\def\nn{\nonumber}
\newcommand{\ls}[1]
   {\dimen0=\fontdimen6\the\font 
    \lineskip=#1\dimen0
    \advance\lineskip.5\fontdimen5\the\font
    \advance\lineskip-\dimen0
    \lineskiplimit=.9\lineskip
    \baselineskip=\lineskip
    \advance\baselineskip\dimen0
    \normallineskip\lineskip
    \normallineskiplimit\lineskiplimit
    \normalbaselineskip\baselineskip
    \ignorespaces}

\def\oldreffmt#1{\rlap{[#1]} \hbox to 2\parindent{}}

\def\figfmt#1{\rlap{Figure {#1}} \hbox to 1in{}}

\def\etal{\hbox{\it et al.}}

\def\VEV#1{\left\langle #1\right\rangle}

\def\slash#1{#1\!\!\!/\!\,\,}
\def\beq{\begin{equation}}
\def\eeq{\end{equation}}
\def\bea{\begin{eqnarray}}
\def\eea{\end{eqnarray}}
\def\half{\frac{1}{2}}

\def\bq{\begin{quote}}
\def\eq{\end{quote}}

\def\half{\frac{1}{2}}

\def \gta {\mathrel{\vcenter
     {\hbox{$>$}\nointerlineskip\hbox{$\sim$}}}}

\def \etal {{\it et al.}\ }

\newcommand{\be}{\begin{equation}}
\newcommand{\ee}{\end{equation}}
\newcommand{\bdm}{\begin{displaymath}}
\newcommand{\edm}{\end{displaymath}}
\def\simlt{\mathrel{\lower2.5pt\vbox{\lineskip=0pt\baselineskip=0pt
           \hbox{$<$}\hbox{$\sim$}}}}
\def\simgt{\mathrel{\lower2.5pt\vbox{\lineskip=0pt\baselineskip=0pt
           \hbox{$>$}\hbox{$\sim$}}}}
\catcode`@=11
\newcount\@tempcntc
\def\@citex[#1]#2{\if@filesw\immediate\write\@auxout{\string\citation{#2}}\fi
  \@tempcnta\z@\@tempcntb\m@ne\def\@citea{}\@cite{\@for\@citeb:=#2\do
    {\@ifundefined
       {b@\@citeb}{\@citeo\@tempcntb\m@ne\@citea\def\@citea{,}{\bf ?}\@warning
       {Citation `\@citeb' on page \thepage \space undefined}}%
    {\setbox\z@\hbox{\global\@tempcntc0\csname b@\@citeb\endcsname\relax}%
     \ifnum\@tempcntc=\z@ \@citeo\@tempcntb\m@ne
       \@citea\def\@citea{,}\hbox{\csname b@\@citeb\endcsname}%
     \else
      \advance\@tempcntb\@ne
      \ifnum\@tempcntb=\@tempcntc
      \else\advance\@tempcntb\m@ne\@citeo
      \@tempcnta\@tempcntc\@tempcntb\@tempcntc\fi\fi}}\@citeo}{#1}}
\def\@citeo{\ifnum\@tempcnta>\@tempcntb\else\@citea\def\@citea{,}%
  \ifnum\@tempcnta=\@tempcntb\the\@tempcnta\else
   {\advance\@tempcnta\@ne\ifnum\@tempcnta=\@tempcntb \else \def\@citea{--}\fi
    \advance\@tempcnta\m@ne\the\@tempcnta\@citea\the\@tempcntb}\fi\fi}
\catcode`@=12

\def\be{\begin{equation}} 
\def\ee{\end{equation}} 
\def\bea{\begin{eqnarray}} 
\def\eea{\end{eqnarray}} 

\begin{document}
\begin{flushright}
{Fermilab-Conf-98/040-T\\
EFI-Preprint-98-04\\
hep-ph/9802216\\
Jan. 29, 1998\\
}
\end{flushright}

\title{ Topcolor and the First Muon Collider\footnote{
Talk presented at the Workshop on Physics at the First Muon Collider
and at the Front End of the Muon Collider}
}
\author{Christopher T. Hill}
\address{Fermi National Accelerator Laboratory\\
P.O. Box 500, Batavia, Illinois, 60510\\
and\\
The Department of Physics and Enrico Fermi Institute\\
The University of Chicago, Chicago, Illinois
}

\maketitle

\begin{abstract}
We describe a class of models of
electroweak symmetry breaking that involve strong dynamics
and top quark condensation.   A new scheme based upon a 
seesaw mechanism appears particularly promising. Various
implications for the first-stage muon collider are discussed.
\end{abstract}

     
     
%
 
\section*{Topcolor~I   }

The
top quark mass may be large because
it is a combination
of a {\em dynamical condensate
component}, $(1-\epsilon)m_t$,
generated by a new strong dynamics \cite{BHL},
together with a small {\em fundamental component},
$\epsilon m_t$, i.e, $\epsilon<<1$, generated by 
something else.  The most obvious ``handle''
on the top quark for new dynamics is the color index. 
Invoking new dynamics involving the top
quark color index leads directly to a
class of Technicolor--like models
incorporating ``Topcolor''.
We expect in such schemes that the new strong dynamics 
occurs primarily in interactions
that involve $\overline{t}t\overline{t}t$,
$\overline{t}t\overline{b}b$, and
$\overline{b}b\overline{b}b$.

In Topcolor~I the dynamics at the $\sim 1$ TeV scale
involves the following structure at the TeV scale
(or a generalization thereof) \cite{TC2}: 
\beq
SU(3)_1\times SU(3)_2
\times U(1)_{Y1}\times U(1)_{Y2}
\times SU(2)_L \rightarrow
SU(3)_{QCD}\times U(1)_{EM}
\eeq
where
$SU(3)_1\times U(1)_{Y1}$
($SU(3)_2\times U(1)_{Y2}$)
generally couples preferentially
to the third (first and
second) generations.  The
$U(1)_{Yi}$
are just strongly rescaled
versions of
electroweak  $U(1)_{Y}$.

The fermions are then assigned
$(SU(3)_1, SU(3)_2, {Y_1}, {Y_2}$) quantum numbers in the following
way:
\bea
(t,b)_L \;\;   &\sim  & (3,1,{1}/{3},0) \qquad \qquad
(t,b)_R \sim \left(3,1,({4}/{3},-{2}/{3}),0\right) \\ \nonumber
(\nu_\tau,\tau)_L &\sim & (1,1,-1,0) \qquad \qquad
\tau_R \sim \left(1,1,-2,0\right) \\ \nonumber
  & & \\ \nonumber
(u,d)_L,\;\;  (c,s)_L & \sim & (1,3,0,{1}/{3}) \qquad \qquad
(u,d)_R, \;\; (c,s)_R \sim \left(1,3,0,({4}/{3},-{2}/{3})\right) \\  
\nonumber
(\nu, \ell)_L\;\; \ell = e,\mu & \sim & (1,1,0,-1) \qquad \qquad
\ell_R \sim \left(1,1,0,-2\right)
\eea
Topcolor must be broken, which we describe by
an (effective) scalar field:
\beq
\Phi \sim  (3,\bar{3}, y, -y) \label{phi_q}
\eeq
When $\Phi$ develops a VEV,
it produces the  simultaneous symmetry breaking
\beq
SU(3)_1\times SU(3)_2 \rightarrow
SU(3)_{QCD}\qquad
\makebox{ and}
\qquad
U(1)_{Y1}\times U(1)_{Y2}
\rightarrow  U(1)_{Y}
\label{sym_bre}
\eeq
$SU(3)_1\times U(1)_{Y1}$ is assumed to be
strong enough to form
chiral condensates which will
be ``tilted'' in the top
quark direction by the $U(1)_{Y1}$ couplings.
The theory is assumed to spontaneously break down to ordinary QCD
$\times U(1)_{Y}$ at a scale of $\sim 1$~TeV, before it becomes confining.
The isospin splitting that permits the formation of a $\VEV{\overline{t}t}$
condensate but disables the $\VEV{\overline{b}b}$ condensate is due to the
$U(1)_{Yi}$ couplings. 
The $b$--quark mass in this scheme
can arise from a combination
of ETC effects and instantons
in  $SU(3)_1$. The $\theta$--term
in $SU(3)_1$ may manifest itself as
the CP--violating phase in the CKM matrix.
Above all, the new spectroscopy
of such a system
should begin to materialize
indirectly
in the third generation,
perhaps at the Tevatron in top
and bottom quark production, or possibly in a muon collider.

The symmetry breaking pattern outlined above will generically give rise to
three (pseudo)--Nambu--Goldstone bosons $\tilde{\pi}^a$,
or``top-pions'', near the top mass scale. {\em This  is  the 
smoking gun of Topcolor.}  [We were led to Topcolor by considering 
how strong dynamics might produce the analog of the 
decay $t\rightarrow H^+ + b$,
considered to be a SUSY signature for a charged Higgs-boson 
$H$.  This is an example of ``SUSY-Technicolor/Topcolor duality''.]
If the Topcolor scale is of the order
of 1~TeV, the top-pions will
have a decay constant
of $f_\pi  \approx 50$ GeV, and a
strong coupling
given by a Goldberger--Treiman
relation,
$g_{tb\pi} \approx m_t/\sqrt{2}f_\pi\approx 2.5$,
potentially
observable in
$\tilde{\pi}^+\rightarrow t + \overline{b}$
if $m_{\tilde{\pi}} > m_t + m_b$.

We assume presently that ESB can be primarily driven
by a Higgs sector or Technicolor, with gauge group $G_{TC}$\cite{Lane} \cite{Lane2}.
This gives the ${\cal{O(\epsilon)}}$ component of $m_t$.
Technicolor can also provide
condensates which generate the
breaking of Topcolor
to QCD and $U(1)_Y$.

The coupling constants (gauge
fields) of
$SU(3)_1\times SU(3)_2$  are
respectively
$h_1$ and $h_2$ ($A^A_{1\mu}$
and $A^A_{2\mu}$)
while for $U(1)_{Y1}\times U(1)_{Y2}$
they
are respectively  ${q}_1$ and $q_2$,
$(B_{1\mu}, B_{2\mu})$.
The $U(1)_{Yi}$ fermion couplings are
then $q_i\frac{Yi}{2}$, where $Y_1, Y_2$
are the charges of the fermions under $U(1)_{Y1}, U(1)_{Y2}$ respectively.

Topcolor I produces new gauge heavy bosons $Z'$, and 
``colorons'' $B^A$ with couplings to fermions
given by:
\begin{equation}\label{lzb}
{\cal L}_{Z'}=g_1(Z'\cdot J_{Z'}) \qquad
{\cal L}_{B}=g_3\cot\theta (B^A\cdot J^A_{B})
\end{equation}
where the currents $J_{Z'}$ and $J_B$ in general involve all three
generations of fermions
\bea\label{jzb123}
J_{Z'} & = & -(J_{Z',1}+J_{Z',2})\tan\theta'  +J_{Z',3}\cot\theta'
\\ \nonumber  
J_B  & = & -(J_{B,1}+J_{B,2})\tan\theta  + J_{B,3}\cot\theta 
\eea
For example, for the third generation the currents read explicitly
(in a weak eigenbasis):
\bea\label{jz3}
J^\mu_{Z',3} &=& \frac{1}{6}\bar t_L\gamma^\mu t_L+
\frac{1}{6}\bar b_L\gamma^\mu b_L+\frac{2}{3}\bar t_R\gamma^\mu t_R
-\frac{1}{3}\bar b_R\gamma^\mu b_R \\ \nonumber
& & -\frac{1}{2}\bar\nu_{\tau L}\gamma^\mu \nu_{\tau L}
-\frac{1}{2}\bar\tau_L\gamma^\mu\tau_L-\bar\tau_R\gamma^\mu\tau_R
\\ \nonumber
J^{A,\mu}_{B,3} & = & \bar t\gamma^\mu\frac{\lambda^A}{2}t+
\bar b\gamma^\mu\frac{\lambda^A}{2}b
\eea
where $\lambda^A$ is a Gell-Mann matrix acting on color indices.
We ultimately demand $\cot\theta \gg 1$
and  $\cot\theta' \gg 1$
to select the top quark direction for condensation.

The attractive Topcolor interaction, for sufficiently large
$\kappa = g^2_3\cot^2\theta/4\pi $, would by itself
trigger the formation of a
low energy
condensate, $\VEV{\overline{t}t + \overline{b}b}$, which
would break $SU(2)_L\times SU(2)_R\times  U(1)_Y
\rightarrow U(1)\times SU(2)_{c}$, where $SU(2)_{c}$ is
a global custodial symmetry. On the
other hand,  the $U(1)_{Y1}$ force is attractive in the  
${\overline{t}t}$
channel and repulsive in the ${\overline{b}b}$ channel.  Thus, 
to make $\VEV{\overline{b}b} = 0 $ and  $\VEV{\overline{t}t}
\neq 0$ we can
have in concert critical and subcritical values of the combinations:
\beq
\kappa + \frac{2\,\kappa_1}{9N_c} > \kappa_{crit} ;
\qquad
\kappa_{crit} > \kappa - \frac{\kappa_1}{9N_c};
\label{crit_con}
\eeq
Here $N_c$ is the number of colors and $\kappa_1 = g_1^2\cot^2\theta'/4\pi $. 
(It should be mentioned that our analyses
are performed in the context of a large-$N_c$ approximation). 
This leads to ``tilted'' gap equations in which the top
quark acquires a constituent mass, while the $b$ quark
remains massless.  Given that both $\kappa$ and $ \kappa_1$
are large there is no particular fine--tuning occuring here,
only ``rough--tuning'' of the desired tilted configuration.
Of course, the NJL approximation is crude, but as long as the  
associated
phase transitions of the real strongly
coupled theory are approximately second order, analogous
rough--tuning in the full theory is possible.
The full phase diagram of the model is shown in Fig.~1.
of \cite{Burdman}.

\section*{Topcolor~II}

If the above described ``Topcolor I" is the analog of Weinberg's
original version of the SM, incorporating standard fermions
and the $Z$-boson,  then Topcolor II is the analog of
the original Georgi-Glashow model, which incorporated no new $Z$ boson,
but rather included additional fermions.  [This is an example of 
``Weinberg---Georgi-Glashow'' duality.]
The strong $U(1)$ is present in the previous scheme to avoid
a degenerate $\VEV{\bar{t}t}$ with $\VEV{\bar{b}b}$.  However,
we can give a model in which there is: (i) a Topcolor $SU(3)$
group but (ii) no strong $U(1)$ with (iii) an anomaly-free
representation content. In fact the original
model of \cite{TC2} was of this form, introducing a
new quark of charge $-1/3$.  Let us consider a
generalization of this scheme which consists of the gauge structure
$SU(3)_Q\times SU(3)_1\times SU(3)_2 \times U(1)_{Y}\times SU(2)_{L}$.
We require an additional triplet of fermions
fields $(Q_R^a)$ transforming as $(3,3,1)$
and $Q_L^{\dot{a}}$ transforming as $(3,1,3)$ under
the $SU(3)_Q\times SU(3)_1\times SU(3)_2$.

The fermions are then assigned the following quantum numbers
in $SU(2)\times SU(3)_Q \times SU(3)_1\times SU(3)_2\times U(1)_Y $:
\bea
(t,b)_L \;\;  (c,s)_L &\sim & (2,1,3,1) \qquad Y=1/3 \\ \nonumber
(t)_R &\sim & (1,1,3,1) \qquad Y=4/3;\qquad \\\nonumber
(Q)_R &\sim  & (1,3,3,1) \qquad Y=0\\ \nonumber
  & & \\ \nonumber
(u,d)_L &\sim & (2,1,1,3) \qquad Y=1/3 \\ \nonumber
(u,d)_R \;\; (c,s)_R &\sim & (1,1,1,3) \qquad Y=(4/3,-2/3) \\  
\nonumber
(\nu, \ell)_L\;\; \ell = e,\mu,\tau &\sim & (2,1,1,1) \qquad Y=-1;  
\qquad
\\ \nonumber (\ell)_R &\sim &  (1,1,1,1) \qquad Y=-2 \\ \nonumber
b_R & \sim & (1,1,1,3) \qquad Y= 2/3;   \qquad
\\ \nonumber  (Q)_L
& \sim & (1,3,1,3) \qquad Y= 0;
\eea
 Thus, the $Q$ fields are electrically
neutral.  One can verify that this assignment is anomaly free.

The $SU(3)_Q$
confines and
forms a $\VEV{\bar{Q}Q}$ condensate which acts
like the $\Phi$ field and breaks the Topcolor
group down to QCD dynamically.    We assume that $Q$ is then  
decoupled from the
low energy spectrum by its large constituent mass.  There is a  
lone
$U(1)$ Nambu--Goldstone boson  $\sim {\bar{Q}\gamma^5 Q}$
which acquires a large mass by $SU(3)_Q$
instantons.

\section*{~Triangular Textures }

The texture of the fermion mass matrices will generally
be controlled by the symmetry breaking pattern of a horizontal
symmetry.  In the present case we are specifying a residual
Topcolor symmetry, presumably  subsequent to some
initial breaking at some 
scale $\Lambda$, large compared to Topcolor, e.g., the third
generation fermions in Model~I have different Topcolor assignments than  
do the
second and first generation fermions. Thus the texture will depend
in some way upon the breaking of Topcolor \cite{Burdman} \cite{Lane}.

Let us study a fundamental Higgs boson, which ultimately
breaks $SU(2)_L\times U(1)_Y$,
together with an effective field $\Phi$  breaking Topcolor
as in eq.(\ref{sym_bre}).  We must now specify the full Topcolor  
charges
of these fields. As an example, under
$SU(3)_1\times SU(3)_2 \times U(1)_{Y1}\times U(1)_{Y2}\times  
SU(2)_L$
let us choose:
\beq
\Phi \sim (3,\bar{3}, \frac{1}{3}, -\frac{1}{3}, 0)
\qquad
H \sim (1,1,0, -1, \half)
\eeq
The effective couplings to fermions that generate mass terms in the
up sector are of the form
\by
{\cal L}_{{\cal M}_U}&=&m_0 \bar{t}_Lt_R +c_{33}\bar{T}_Lt_R  
H\fr{\det\F^\dagger}{\L^3}+
c_{32}\bar{T}_L c_R H\fr{\F}{\L} + c_{31}\bar{T}_L u_R  
H\fr{\F}{\L}\nn\\
& &+c_{23}\bar{C}_L t_R H \F^\dagger  
\fr{\det\F^\dagger}{\L^4}+c_{22}\bar{C}_L c_R
H + c_{21}\bar{C}_L u_R H  \label{lag_mass}\\
& & + c_{13}\bar{F}_L t_R H \F^\dagger  
\fr{\det\F^\dagger}{\L^4}+c_{12}\bar{F}_L c_RH
+c_{11}\bar{F}_L u_R H  + {\rm h.c.} \nn
\ey
Here $T=(t,b)$, $C=(c,s)$ and $F=(u,d)$.  The mass $m_0$ is
the dynamical condensate top mass.
Furthermore $\det\Phi$ is defined by
\bz
\det \Phi \equiv \frac{1}{6}\e_{ijk}\e_{lmn}\Phi_{il}\Phi_{jm}\Phi_{kn}
\label{phi_det}
\ez
where in $\Phi_{rs}$ the first(second) index refers to $SU(3)_1$ ($SU(3)_2$).  
The matrix elements now require factors
of $\Phi$ to connect the third with the
first or second generation color indices. The down quark
and lepton mass matrices are generated by couplings analogous to
(\ref{lag_mass}).

To see what kinds of textures can arise naturally,
let us assume that the ratio $\Phi/\Lambda$ is small, O($\epsilon$).
The field $H$ acquires a VEV of $v$.
Then the resulting mass  matrix is approximately triangular:
\bea
&& \left( \begin{array}{ccc}
c_{11}v &  c_{12}v & \sim 0    \cr 
 c_{21}v & c_{22}v  &\sim 0    \cr
c_{31}O(\epsilon)v & c_{32}O(\epsilon)v & \sim m_0 + O(\epsilon^3)v    
\cr
 \end{array}\right)\label{m_trian}
\eea
where we have kept only terms of $\cal O (\e)$ or larger.

This is a triangular matrix (up to the $c_{12}$ term).  
When it is written in the form
$U_L {\cal D} U^{\dagger}_R$ with $U_L$ and $U_R$ unitary and ${\cal  
D}$
positive diagonal,
there automatically result restrictions on $U_L$ and $U_R$.  In the  
present case,
the elements $U^{3,i}_L$ and $U^{i,3}_L$ are vanishing for
$i\neq 3$ , while the elements of $U_R$ are not constrained
by triangularity.
Analogously, in the down quark sector $D^{i,3}_L=D^{3,i}_L=0$ for  
$i\neq 3$
with $D_{R}$ unrestricted.
The situation is reversed when the opposite corner elements are  
small,
which can be achieved by choosing $H \sim (1,1, -1, 0, \half)$.

These restrictions on the quark mass rotation matrices have important
phenomenological consequences.
For instance, in the process $B^0\rightarrow \overline{B^0}$ there are  
potentially
large contributions from top-pion and coloron exchange.  However, these contributions are proportional to the
product $D^{3,2}_L D^{3,2}_R$.  The same occurs in $D^0-\bar{D}^0$  
mixing, where
the effect goes as products involving $U_L$ and $U_R$ off-diagonal  
elements.
Therefore, triangularity can naturally select
these products to be small.

The precise selection rules depend upon the particular symmetry  
breaking that occurs. This example is merely illustrative of the
systematic effects that can occur in such schemes.

\section*{Top-pions; Instantons; The b-quark mass.}
Since the 
top condensation is a spectator 
to the TC (or Higgs) driven
ESB, there must
occur a multiplet of top-pions.
A chiral Lagrangian can be written:
\beq
L = i\overline{\psi}\slash{\partial}\psi - m_t(
\overline{\psi}_L\Sigma P\psi_R + h.c.) -\epsilon m_t
\overline{\psi}P\psi, \qquad
P=\left(\begin{array}{cc} 1 & 0\\ 0 & 0
\end{array}\right)
\eeq
and $\psi=(t,b)$, and $\Sigma = \exp(i\tilde{\pi}^a\tau^a
/\sqrt{2}f_\pi)$. With $\epsilon = 0$ this is invariant under
$\psi_L\rightarrow e^{i\theta^a\tau^a/2}\psi_L$,
$\tilde{\pi}^a\rightarrow \tilde{\pi}^a + 
\theta^a f_\pi/\sqrt{2}$.
Hence, the relevant currents are left-handed,
$j_\mu^a = \psi_L\gamma_\mu\frac{\tau^a}{2}\psi_L$,
and $<\tilde{\pi}^a|j_\mu^b|0> 
= \frac{f_\pi}{\sqrt{2}}p_\mu 
\delta^{ab}$.  The Pagels-Stokar relation, eq.(1),
then follows by demanding that the 
$\tilde{\pi}^a$ kinetic
term is generated by integrating 
out the fermions.  The 
top--pion decay constant estimated 
from eq.(1) using
$\Lambda = M_{B}$ and $m_t = 175$ 
GeV is $f_\pi \approx 50$ GeV.
The couplings of the top-pions 
take the form:
\beq
 \frac{m_t}{\sqrt{2}f_\pi} 
\left[ {i}\overline{t} 
\gamma^5 t \tilde{\pi}^0 
+\frac{i}{\sqrt{2} }
\overline{t} (1-\gamma^5) b \tilde{\pi}^+ 
+ \frac{i}{\sqrt{2} }
\overline{b} (1+\gamma^5)t  \tilde{\pi}^- 
\right]
\eeq
and the coupling strength is 
governed by the relation 
$g_{bt\tilde{\pi}} \approx m_t/\sqrt{2}f_\pi$.

The small ETC mass component of 
the top quark implies that the masses
of the top-pions will depend 
upon $\epsilon$ and $\Lambda$. 
Estimating the induced top-pion 
mass from the fermion loop yields:
\beq
m_{\tilde{\pi}}^2 = 
\frac{N \epsilon m_t^2 M_B^2 }{8\pi^2 f_\pi^2} 
= \frac{\epsilon M_B^2 }{\log(M_B/m_t)}
\eeq
where the Pagels-Stokar formula 
is used for $f_\pi^2$
(with $k=0$) in the
last expression. For 
$\epsilon = (0.03,\; 0.1)$, 
$M_B\approx (1.5,\; 1.0) $ TeV, 
and $m_t=180$ GeV  this 
predicts $m_{\tilde{\pi}}= (180,\; 240)$ GeV.
The bare value of $\epsilon$ 
generated at the  ETC scale 
$\Lambda_{ETC}$, however, 
is subject to very large 
radiative enhancements by 
Topcolor and $U(1)_{Y1}$ by 
factors of order 
$(\Lambda_{ETC}/M_B)^p \sim 10^1$,
where the $p\sim O(1)$. 
Thus, we expect that
even a  bare value of 
$\epsilon_0 \sim 0.005$
can produce sizeable $m_{\tilde{\pi}} > m_t$.
Note that $\tilde{\pi}$ will generally 
receive gauge contributions 
to it's mass; these are
at most electroweak in strength, 
and therefore of
order $\sim 10$ GeV.

Top-pions can be as light as $\sim 150$ GeV, in
which case they would emerge as a detectable branching
fraction of top decay \cite{BB}.  However, there are
dangerous effects in $Z\rightarrow b\bar{b}$ with
low mass top pions and decay constamnts as small as $\sim
60 $ GeV \cite{burdman2}.  A more comfortable 
phenomenological range is slightly larger
than our estimates, $m_{\tilde{\pi}} \gta 300$ GeV and
$f_\pi \gta 100$ GeV.

The $b$ quark receives 
mass contributions from ETC of $O(1)$
GeV, but also an induced mass
from instantons in $SU(3)_{1}$. 
The instanton
effective Lagrangian may be 
approximated by the
`t Hooft flavor determinant (
we place the cut-off at $M_B$):
\beq
L_{eff} = \frac{k}{M_B^2} 
e^{i\theta_1} \det 
(\overline{q}_L q_R) + h.c.
= \frac{k}{M_B^2} 
e^{i\theta_1}[(\overline{b}_L b_R)
(\overline{t}_Lt_R)
- (\overline{t}_L b_R)
(\overline{b}_Lt_R)] + h.c.
\eeq
where $\theta_1$ is the 
$SU(3)_1$
strong $CP$--violation 
phase. $\theta_1$ 
cannot be eliminated 
because
of the ETC contribution 
to the $t$ and $b$ masses.  
It 
can lead to
induced scalar couplings 
of the neutral top--pion \cite{Burdman},
and
an induced CKM CP--phase, 
however, we will presently
neglect the effects of 
$\theta_1$.

We generally expect $k\sim 1$ 
to $10^{-1}$ as in QCD.
Bosonizing in fermion bubble 
approximation
$\overline{q}^i_Lt_R \sim 
\frac{N}{8\pi^2} m_t M_B^2 
\Sigma^i_1$, where 
$\Sigma^i_j = 
\exp(i\tilde{\pi}^a\tau^a/\sqrt{2}f_\pi)^i_j$ yields:  
\beq
L_{eff} \rightarrow  
\frac{Nk m_t}{8\pi^2}  
e^{i\theta}[(\overline{b}_L b_R)
\Sigma^1_{1} +
(\overline{t}_Lb_R)
\Sigma^2_{1} + h.c.]
\eeq
This implies an instanton 
induced $b$-quark mass:
\beq
m^\star_b \approx 
\frac{3 k m_t}{8\pi^2} \sim 6.6\; k\; GeV 
\eeq
This is not an unreasonable 
estimate of the observed $b$ quark mass,
as we might have feared it
 would be too large. 

\newcommand{\bear}{\begin{eqnarray}}
\newcommand{\eear}{\end{eqnarray}}
\newcommand{\lae}{\begin{array}{c}\,\sim\vspace{-21pt}\\< \end{array}}
\newcommand{\gae}{\begin{array}{c}\,\sim\vspace{-21pt}\\> \end{array}}
\newcommand{\inl}{{\scriptscriptstyle L}}
\newcommand{\inr}{{\scriptscriptstyle R}}
\def\vbr{$\vphantom{\sqrt{F_e^i}}$}
\def\VEV#1{\left\langle #1\right\rangle}

\section*{Top See-Saw}

 EWSB may
occur via the condensation of the top quark 
in the presence of an extra vectorlike, weak-isoscalar quark \cite{Bogdan}. 
The mass scale of the condensate is large, of order 0.6 TeV
corresponding to the electroweak scale $f_{\pi} \approx 175$ GeV.
The vectorlike iso-scalar then naturally admits
a seesaw mechanism, yielding the physical top quark mass, which is then
adjusted to the experimental value.  The choice of a natural $\sim $TeV scale
for the topcolor dynamics then determines the mass of the weak-isoscalar
see-saw partner. 
The scheme is economical, requiring no additional
weak--isodoublets, and therefore
easily satisfies the constraints
upon the $S$ parameter using estimates made in the large--N approximation.
 The constraints 
on custodial symmetry violation, i.e., the value of the 
$\delta \rho$ or equivalently, $T$ parameter, are easily satisfied,
being principally the usual $m_t$ contribution, plus corrections that
are suppressed by the see-saw mechanism.

The dynamical fermion masses that are induced
can be written as:
\be
{\cal L} = - \left( \overline{t}_L \ , \ \overline{\chi}_L  \right) 
\left( \ba{rl} 0 \;\;\;  & m_{t \chi} \\ 
        m_{\chi t} & m_{\chi \chi} \ea \right) 
\left( \ba{c} t_R \\ \chi_R \ea \right) + {\rm h.c.} 
\label{massterm}
\ee
Typically
$\overline{\chi}_L \chi_R$ is
the most attractive channel,  and it is possible to arrange the 
$\langle \overline{\chi}_L \chi_R \rangle$ condensate to be significantly
 larger than the other ones, such that
$ m_{\chi \chi}^2 \gg  m_{\chi t}^2 >  m_{t \chi}^2 ~$.
As a result the physical top mass is suppressed by a seesaw mechanism:
\be
m_t \approx \frac{m_{\chi t} m_{t \chi}}{  m_{\chi \chi}} \left[
1 + O\left(m_{\chi t, \, t\chi}^2/m_{\chi \chi}^2\right) \right]~.
\label{topmass}
\ee
The electroweak symmetry is broken by the $m_{t \chi}$
dynamical mass. Therefore, the electroweak scale is estimated to be
given by
\be
v^2 \approx \frac{3}{16 \pi^2} m_{t\chi}^2 
\ln \left(\frac{M}{m_{t\chi}}\right) ~.
\ee
Thus, $v \approx 174$ GeV requires a dynamical mass
$m_{t \chi} \sim 620 \,{\rm GeV}  $
for $M \sim 5$ TeV (and $m_{t \chi} = 520$ GeV for $M \sim 10$ TeV).
From eq.~(\ref{topmass})
follows then that a top mass of 175 GeV requires
$ {m_{\chi t}}/{  m_{\chi \chi}} \approx 0.29 ~$.
The electroweak
$T$ parameter 
can be estimated in fermion--bubble large--$N$ approximation as: 
\be
T \approx \frac{3 m_t^2}{16 \pi^2 \alpha(M_Z^2) v^2} 
\frac{m^2_{t \chi}}{m^2_{\chi t}} \left[
1 + O\left(m_{\chi t, \, t\chi}^2/m_{\chi \chi}^2\right) \right] ~,
\ee
where $\alpha$ is the fine structure constant.
Moreover, we obtain the usual Standard Model result
for the $S$ parameter.
Requiring that our model does not exceed the 
1$\sigma$ upper bound on $S$ and $T$, we obtain 
$ {m_{t \chi}}/{m_{\chi t}} \le 0.55 ~$.

It should be emphasized that these results 
do not require excessive fine-tuning.  The top-seesaw
is therefore a plausible natural theory of dynamical EWSB
with a minimal number of new degrees of freedom.
This model also implies the existence of 
pseudo--Nambu-Goldstone bosons (pNGB's).  A cursory
discussion of that is given in ref.\cite{Bogdan}.

\section*{ Observables }

There are several classes of possible experimental implications of the
kinds of models we described above that may be relevant to the muon collider.
We will describe them here briefly as lines
to be developed further. These may be enumerated as follows:

\begin{enumerate}
\item $\mu \overline{\mu } \rightarrow Z'$; this is the province of 
high energy machine, since we expect $M_{Z'} \gta 0.5 $ TeV. 

\item $\mu \overline{\mu } \rightarrow \pi_{top} $; the notion that the muon collider can see technipions, or other PNGB's, such as top-pions has emerged from discussions in this workshop, prompted by MacKenzie and myself.
Lane has presented the multi-scale 
technicolor signal \cite{Lane2}. 

\item Effects in $Z$ physics involving the third generation, such as
$Z\rightarrow b\bar{b}$ \cite{burdman2}.

\item Effects in top-quark pair production at threshold, e.g.,
see \cite{Peskin} for analogous case in $e^+e^-$ and
$p\overline{p}$ collider physics.

\item Induced GIM violation in low energy processes such as $K^+\to \pi^+ \nu\bar{\nu}$; we discuss this below as an example of a
potential signature that can be enhanced by Topcolor wrt the Standard Model (this result was anticipated in ref\cite{Burdman} before the observation of the single event at Brookhaven E787).

\item Induced lepton family number violation, e.g. $\mu \overline{\mu } \rightarrow \tau \overline{\mu}$.

\item Flavor dependent production effects, e.g. anomalous $\mu \overline{\mu } \rightarrow b\overline{s}$, etc.

\item New physics in e.g. $\mu p$ collisions, such as  $d(u) + \overline{\mu } \rightarrow b(t) +\overline{\tau}$.
\end{enumerate}

GIM and lepton family number violation arise because of the generational
structure of topcolor.  (It is actually more general than topcolor; the
mere statement than the top mass is largely dynamical implies effects
like this)
In going to the mass eigenbasis, quark (and lepton) fields are rotated,
e.g.,  by the matrices
$U_L$, $U_R$ (for the up-type left and right handed quarks) and $D_L$, $D_R$
(for the down-type left and right handed quarks).
For example, for the b-quark we make the replacement
\bz
b_L\to D_{L}^{bb}b_L + D_{L}^{bs}s_L + D_{L}^{bd}d_L \label{bl_rot}
\ez
and analogously for $b_R$. Thus there will be induced FCNC interactions.
This provides constraints and opportunities.
Thus, induced
effects like $\mu \overline{\mu } \rightarrow b\overline{s}$
may be enhanced, and effects like $\mu \overline{\mu } \rightarrow \tau \overline{\mu}$ may occur.  
Since the muon is presumably closer 
in affiliation to the third generation than is
the electron, such effects may show up in muon collider physics, but be
inaccessible in electron linear colliders! Similarly, induced
effects like $\mu \overline{\mu } \rightarrow b\overline{s}$
may be enhanced.  

For the FMC, sensitive probes arise
in e.g., K-physics. there is a $ Z'$ induced contact term at low energies of the form $b\overline{b}\nu_\tau \overline{\nu}_\tau$ (this assumes that the $\tau$ is associated with the third generation; nothing fundamentally compels this,
but we shall assume it to be true in the following).
The above mass rotation induces a  $s\overline{d}\nu_\tau \overline{\nu}_\tau$
which contributes to  $K^+\to \pi^+ \nu\bar{\nu}$.
The ratio of the Topcolor amplitude to the SM is then
\bz
\frac{{\cal A}^{TC}}{{\cal A}^{SM}}=-\left(\frac{g_1\cot\q
'}{M_{Z'}}\right)^2\;\frac{\sqrt{2}\pi\sin^2\theta_W}{24\a\; G_F}\;
\frac{\d_{ds}}{\sum_{j}V^{*}_{js}V_{jd} D_j(x_j)} 
\sim
-3\times 10^9\; \d_{ds}\;
\frac{\kappa_1}{M_{Z'}^{2}} \label{ratio}
\ez
where $\d_{ds}^{*}=D_{L}^{bs}D_{L}^{bd*}-2D_{R}^{bs}D_{R}^{bd*}$. The
form-factor $f_+(q^2)$ is experimentally well known. 
We expect, $|\d_{ds}| \sim \lambda^{10}\;$
where $\lambda$ is the Wolfenstein CKM parameter.
For $M_{Z'}=500$~GeV and  $\kappa_1=1$
the ratio of amplitudes is about $\sim 4.0$, and the 
branching ratio is between $0.3$ to $O(10)$, times the SM result,
depending on the sign of the interference.
The recent observation of one
event by the Brookhaven E787 Collaboration \cite{BNL} 
makes this an exciting channel
in which to search for new physics.  High sensitivity experiments are 
possible at the front-end  muon collider with its copious K-meson yields.



\end{document}

\bibitem{RGFP} B. Pendleton, G.G. Ross, {\em Phys. Lett.}
{\bf 98B} 291, (1981); C.T. Hill,
{\em Phys. Rev.} {\bf D24}, 691 (1981); C.~T. Hill, C.~N. Leung, 
S. Rao, {\em Nucl. Phys.} {\bf B262},  517 (1985).

\bibitem{Carena} see, for example,  W. A. Bardeen, M. Carena, S. Pokorski, 
C. E. M. Wagner,
{\em Phys. Lett. } {\bf B320}, 110, (1994). 

\bibitem{Miransky} R.R. Mendel, V.A. Miranskii, {\em Phys. Lett.} {\bf B268} 384, (1991); V.A. Miranskii, {\em  Int. J. Mod. Phys.} {\bf A6}
1641, (1991). 

\bibitem{TC} S. Weinberg, {\em Phys. Rev.}
{\bf D13}, 974 (1976);
L. Susskind, {\em Phys. Rev.}
{\bf D20}, 2619 (1979);
 S. Dimopoulos, L. Susskind,
{\em  Nucl. Phys.} {\bf B155}
237 (1979); E. Eichten, K. Lane,
{\em  Phys. Lett.} {\bf 90B}
125 (1980).


\bibitem{Ap} T.~Appelquist,
M.~Einhorn, T.~Takeuchi,
L.~C.~R.~Wijewardhana, {\em  Phys. Lett.} {\bf B220}
223 (1989).  T. W. Appelquist,
D. Karabali, L.C.R. Wijewardhana
Phys. Rev. Lett. {\bf 57}, 957 (1986);
R. R. Mendel, V. Miransky,
{\em Phys. Lett.} {\bf B268}
384 (1991);  V. Miransky,
{\em Phys. Rev. Lett.} {\bf 69}
1022 (1992); N. Evans,
{\em Phys. Lett.} {\bf B331} 378 (1994).

\bibitem{WETC} B. Holdom,
{\em  Phys. Rev. Lett.} {\bf 60}
1223 (1988); see also \cite{Ap}.

\bibitem{Schizon} see C.~T. Hill,
G. G. Ross,
{\it Nucl. Phys.} {\bf B311}, 253 (1988);
{\it Phys. Lett.}, {\bf B203}, 125 (1988),
for a discussion of analogous chiral Lagrangians
and the effects of $CP$--violation.

\bibitem{Hew}  See, e.g., N. Deshpande in
``B-Decays,'' ed. S. Stone,
World Scientific, (1992);  J. L. Hewett,
``Top Ten Models Constrained By
$b\rightarrow s\gamma$, SLAC-PUB-6521
(1994); {\em Phys. Rev. Lett.},
{\bf 70} 1045 (1993); V. Barger,
M. Berger, R.J.N. Phillips,
{\em Phys. Rev. Lett.}, {\bf 70}
1368 (1993); B. Grinstein, R. Springer and M. Wise,
{\em Nucl. Phys.} {\bf B339} 269 (1990).

\bibitem{EL} E. Eichten, K. Lane,
{\em Phys. Lett.} {\bf B222} 274 (1989);
K. Lane, M. Ramana,  {\em Phys. Rev.
}  {\bf D44}, 2678 (1991) .

\bibitem{Chiv} S. Chivukula,
S. Selipsky, E. Simmons,
 {\em Phys. Rev. Lett.}
{\bf 69}, 575  (1992).

\end{document}